\documentclass[10pt]{article} 

\pdfoutput=1
\usepackage{cite}
\usepackage{amssymb}
\usepackage{amsmath}
\usepackage{rawfonts}
\usepackage{graphicx}
\usepackage[usenames,dvipsnames]{color}
\usepackage{bbm}
\usepackage{latexsym}
\usepackage{multirow}
\usepackage{rotating}
\usepackage{lscape}
\usepackage{graphicx} 
\usepackage{subfigure}
\usepackage{xcolor}
\usepackage{fancybox}

\input prepictex
\input pictex
\input postpictex

\usepackage{cmbright}
\usepackage[T1]{fontenc}
\usepackage{graphicx}

\def\q{\quad}

\def\IntN{{\mathbb Z}}

\def\RealN{{\mathbb R}}

\def\mathL{{\mathbb L}}

\def\o#1{\overline{#1}}
\def\w#1{\widetilde{#1}}
\def\Ref#1{(\ref{#1})}

\def\gammas#1{{\gamma_#1}}

\def\sfrac#1#2{\hbox{\nor $\frac{#1}{#2}$}}

\def\L{\left(}         \def\R{\right)}
        
\def\LA{\left\langle}        \def\RA{\right\rangle}

\def\nor{\normalsize}

\def\svv{\,\hbox{$|$}\,}

\def\edge#1#2{{#1{\sim}#2}}

\title{The entropic pressure of lattice knots}
\author{E.J. Janse van Rensburg\\
Department of Mathematics and Statistics, York University \\
Toronto, Ontario, M3J 1P3, Canada}
\date{\today} 

\begin{document}
\maketitle

\begin{abstract}
The entropic pressure in the vicinity of a cubic lattice knot is
examined as a model of the entropic pressure near a knotted
ring polymer in a good solvent.   A model for the scaling of the
pressure is developed and this is tested numerically by sampling
lattice knots using a Monte Carlo algorithm.  Good agreement
is found between scaling predictions and numerical experiments.
\end{abstract}

\section{Introduction}
\label{section1}   

A polymer placed close to a geometric obstacle (such as a hard wall)
loses conformational entropy and this induces a net force on the 
obstacle \cite{MN91}.  This induced force is repulsive and exerts
an average net pressure on the wall.  Entropically induced forces
have been observed experimentally \cite{BDD95,CS95},
and have been simulated using self-avoiding walk models of a 
grafted polymer \cite{DJ13}. 

\begin{figure}[h!]
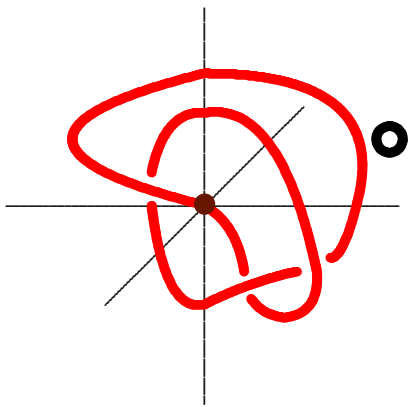

\input figure1.tex
\caption{Schematic illustration of a test particle near a ring 
polymer in three dimensions.  The particle is an obstacle which limits
the conformational degrees of freedom of the polymer.  This reduction
in polymer entropy induces a repulsive net force on the particle.
This force may be described as the gradient of an average 
pressure field in the vicinity of the polymer.}
\label{figure1}    
\end{figure}

A slightly different situation is encountered when a (small)
test particle is placed close to a polymer -- see figure \ref{figure1} 
for a schematic illustration.  The volume occupied by the particle
excludes volume, and this causes a loss of conformational entropy 
in the polymer.  The consequence is that the particle experiences a net 
repulsive force -- if it is free to move, then it will be
expelled from the vicinity of the polymer.  An effect like this
underlies the polymeric stabilisation of a colloid \cite{P91}.   

The forces induced on a test particle near a polymer may be
described as a gradient of a pressure field in the vicinity of
the polymer.  This was examined in several
studies (see for example references \cite{BMJ00,GWF06,MTW77},
and reference \cite{BORW05} for a directed lattice path model).

In reference \cite{GJvR13} a two dimensional self-avoiding walk model
of a ring polymer (a square lattice polygon \cite{H61})
was used to simulate the entropically induced pressure near 
a two dimensional lattice polygon model of a ring polymer.   
A scaling analysis for the net pressure was done and used to 
determine the pressure as a function of distance from the polygon.  

In particular, let $R = \LA r \RA_n$ be a length scale in the model
(such as the root mean square radius of gyration or mean span of a 
polygon of length $n$). Then $R\sim n^\nu$ where $\nu$ is the metric 
exponent of the self-avoiding walk.  In two dimensions the pressure $\mathbf{P}_n(a)$ at 
a distance $aR$ from the polygon, for $a> 0$, was shown to scale as
\begin{equation}
\mathbf{P}_n(a) \simeq C\, g(a)\,a^{-2d}\,n^{2\gammas1 - 1-4\nu - \alpha_s + 2}
\label{eqn1}  
\end{equation}
where $\gammas1$ is the entropic exponent of self-avoiding
walks in a half-space \cite{HG94}, $\alpha_s$ is the entropic exponent
of lattice polygons \cite{D89C}, and $g(a)$ is a scaling function of the
general form
\begin{equation}
g(a) = \int_0^1 x^{\gammas1-1}(1{-}x)^{\gammas1-1}\, f(a/x^\nu)\,
f(a/(1{-}x)^\nu)\,dx .
\end{equation}
The function $f(x)$ is a quickly decaying function (the choice
$f(x) = e^{-x}$ proved consistent with numerical data).

Using the exact values in $d=2$ dimensions in equation \Ref{eqn1} gives
\begin{equation}
n^{19/32} \,\mathbf{P}_n(a) \simeq C\,g(a)\,a^{-4} .
\label{eqn3}   
\end{equation}
This result gave excellent agreement with numerical data \cite{GJvR13}.  

In this paper the pressure near a lattice model of a ring polymer in 
three dimensions is examined.   The model is illustrated schematically in 
figure \ref{figure1} -- a particle is placed near a lattice polygon to measure the 
pressure which may scale in a way similar to the two dimensional result in 
equation \Ref{eqn1}. Substitution of estimated values for exponents in $d=3$
dimensions gives $n^{0.193}P_n(a) \simeq C\,g(a)\,a^{-6}$.    Preliminary 
simulations of three dimensional lattice polygons
showed that this prediction is inconsistent with numerical data.  

In other words, the scaling relation for square lattice polygons in
equation \Ref{eqn1} must be modified for cubic lattice polygons.  This
is done in section \ref{section2},  and the resulting relation is tested
numerically in section \ref{section3}.  The rescaled pressure
${\mathbf P}_n(a)$ for lattice polygons in $d\geq3$ dimensions
is shown to be 
\begin{equation}
\mathbf{P}_n(a) \simeq C
\, n^{\gamma-2d\nu-\alpha_s + 2}\,g_d(a)\,a^{-2d} \;
\hbox{in $d\geq 3$ dimensions,}
\label{eqn6aa}  
\end{equation}
where $g_d(a)$ is given by
\begin{equation}
g_d(a) = \int_0^1 G(a/x^\nu)\,G(a/(1{-}x)^\nu) \,dx .
\end{equation}
As in the definition of $g(a)$ above, the function
$G(x)$ is a quickly decaying function, and choosing an exponential
works well.  More properly, results in references
\cite{dC74,dC80A} suggest that for large values
of $x$ it should be expected that $G(x) \sim x^d e^{-x^\delta}$ 
where $\delta = 1/(1{-}\nu)$; see section \ref{section21}.  

The hyperscaling relation $2{-}\alpha_s = d\nu$ for polygons
in low dimensions can be used to obtain 
\begin{equation}
n^{d\nu - \gamma}\,\mathbf{P}_n(a) \simeq  C\,g_d(a)\,a^{-6}\;
\hbox{if $d= 3$}.
\label{eqn6}  
\end{equation}
For $d>4$ the hyperscaling relation breaks down, and
the relation in equation \Ref{eqn6aa} should be used instead.

Substituting three dimensional estimated values for $\gamma$ and $\nu$
\cite{SBB11,C10} in the above gives $3\nu {-} \gamma \approx
0.606$ so that
\begin{equation}
n^{0.606} \, \mathbf{P}_n(a) \approx  C\,g_d(a)\,a^{-6}\;
\hbox{in $d=3$ dimensions}.
\label{eqn4}  
\end{equation}
This result is quite different from the prediction in equation \Ref{eqn1}.
In section \ref{section2} it will be seen why a different result is 
obtained in $3$ dimensions, compared to the result derived 
for $2$ dimensions in reference \cite{GJvR13}.

In section \ref{section2} the derivation of the scaling relation 
for $\mathbf{P}_n(a)$ in equation \Ref{eqn6} is presented for lattice
polygons in the hypercubic lattice (with $d\geq 3$).  
This derivation is similar to that of equation \Ref{eqn1}, 
but differs in a few important ways due to the 
different dimensionality.  Equation \Ref{eqn1} was 
derived in the square lattice where steric interactions of 
the polygon with itself is an important factor.  A different set of
constraints (or rather the lack thereof) in the cubic lattice (or higher
dimenionsonal lattices) gives a different result if $d\geq 3$.

Since lattice polygons in the cubic lattice are embeddings of the
circle in three space, they have well defined knot types.  Cubic lattice
polygons are \textit{lattice knots}, and they are models of
knotted ring polymer entropy.  In section \ref{section3} it is
argued that the scaling analysis of section \ref{section2} is valid for 
rooted lattice knots as well.   This is tested in section \ref{section3}
by Monte Carlo simulations of lattice knots.  The sampling of 
lattice knots was done by using an implementation of the 
GAS-algorithm for lattice polygons in the cubic lattice
\cite{JvRR09,JvR10} (implemented with BFACF elementary moves \cite{AA83,BF81}).  
Data were collected in several different simulations for
lattice knots of types $K\in\{ 0_1,3_1,4_1,5_1,5_2\}$  (where $0_1$ is the
\textit{unknot}, $3_1$ is the \textit{trefoil}, and $4_1$ is the \textit{figure eight knot}).
For each knot type the calculated pressure should have
scaling given by equation \Ref{eqn4}.  The numerical data confirm this
expectation.  For example, the data show that the pressure is 
similar for different knot types, deviating only at
long distances from one another. In each case the estimated pressure
exhibits scaling consistent with equation \Ref{eqn4}.

\begin{figure}[t!]
\centering
\includegraphics[width=0.55\textwidth,trim=0 10 25 25]{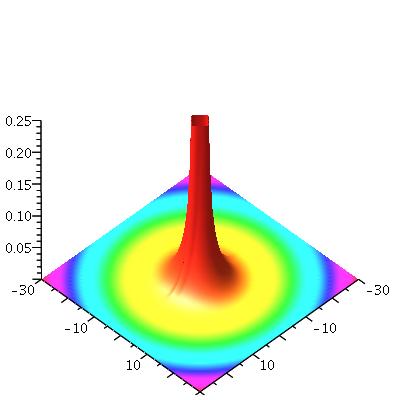}
\caption{The pressure field near a lattice knot of type unknot.
The polygons are rooted at the origin and are of length $100$.
The units on the horizontal axes are in lattice steps.
The pressure peaks sharply towards the origin, and decays to zero 
with increasing distance.  It is isotropic around the origin up to small
corrections close to the origin.}
\label{figure2}    
\end{figure}

The numerical data also show that the pressure field near a rooted
lattice knot is isotropic (independent of direction from the
origin). This result is consistent with the results in reference \cite{GJvR13}.
For example, in figure \ref{figure2} the calculated pressure field in 
the $xy$-plane near a lattice unknot is plotted for polygons of 
length $n=100$.  The pressure increases sharply
as the origin is approached (since the polygon is rooted at the
origin, the pressure at this point is undefined).   The calculated and
interpolated pressure field, even at small distances from the 
origin, is isotropic within numerical error (and with some residual 
effects due to the underlying lattice).  This result is also supported by the 
rescaled data, as will be shown.

The paper is concluded with a few final observations in
section \ref{section4}.

\section{Scaling of the entropic pressure}
\label{section2}   

Denote the standard basis of $\RealN^3$ by $(\vec{e}_1,\vec{e}_2,\vec{e}_3)$.  
Denote the cubic lattice by $\mathL^3$ and let its origin be $\vec{0}$. 
Then $\mathL^3$ is a graph with vertices $\vec{v}\in\IntN^3$ and edges
$\edge{\vec{v}}{\vec{w}}$ joining vertices $\vec{v}$ and $\vec{w}$ if
$\|\vec{v} {-} \vec{w}\|_2=1$.  

A \textit{self-avoiding walk} of length $n$ 
from the origin in $\mathL^3$ is a sequence of 
edges $\LA \edge{\vec{v}_i}{\vec{v}_{i+1}}\svv \hbox{for $i=0,1,\ldots,n{-}1$}\RA$,
where the $\vec{v}_i$ are distinct and $\vec{v}_0 = \vec{0}$.  Walks
are oriented from $\vec{0}$. 
The number of self-avoiding walks of length $n$ edges from $\vec{0}$
is a function denoted by $c_n$.  It is known that the limit
\begin{equation}
\lim_{n\to\infty} \sfrac{1}{n} \log c_n = \log \mu
\label{eqn8aa}  
\end{equation} 
exists (see reference \cite{HM54}), and $\kappa = \log \mu$ is the
\textit{connective constant} of the self-avoiding walk.   
This shows that $c_n = \mu^{n+o(n)}$ -- and $\mu$ is called
the \textit{growth constant}.

A self-avoiding walk $\omega = \LA \edge{\vec{v}_0}{\vec{v}_1},
 \edge{\vec{v}_1}{\vec{v}_2}, \ldots,\edge{\vec{v}_{n-1}}{\vec{v}_n}\RA$
is a \textit{lattice polygon} if $\vec{v}_0 = \vec{v}_n$.  Notice
that lattice polygons are not oriented (by convention).  Two lattice 
polygons are equivalent if the first is a translation of the second
in $\mathL^d$.  Let $p_n$ be the number of distinct lattice polygons 
of length $n$.  For example, if $d=3$, then it may be checked 
that $p_4 = 3$, $p_6=22$ and $p_8=207$.   Since $\mathL^d$ 
is bipartite it follows that $p_{2n+1}=0$. It is known that the limit
\begin{equation}
\lim_{n\to\infty} \sfrac{1}{n} \log p_n = \log \mu
\end{equation}
exists in $\mathL^d$ (if the limit is taken through even values of $n$), 
and $\mu$ is the growth constant of self-avoiding walks defined in
equation \Ref{eqn8aa} \cite{H61,HW62}.  Lattice polygons in $\mathL^d$ are 
models of ring polymer entropy \cite{H61}. 

A lattice polygon $\omega$ is rooted at $\vec{0}$ in $\mathL^d$ if it
contains the origin.  The number of rooted polygons in $\mathL^d$ is
$\o{p}_n = n\,p_n$, since there are $n$ vertices in $\omega$ which can
be placed at the origin.   

The number of self-avoiding walks has asymptotic behaviour
\begin{equation}
c_n \simeq A\,n^{\gamma-1}\,\mu^n
\label{eqnC1}   
\end{equation}
where $\gamma$ is the \textit{entropic exponent} of the self-avoiding walk.
The number of lattice polygons of length $n$ has asymptotic 
behaviour
\begin{equation}
p_n \simeq B\,n^{\alpha_s-3}\,\mu^n ,
\label{eqnP1}   
\end{equation}
where $\alpha_s$ is the entropic exponent of lattice polygons.  This
shows that $\o{p}_n \simeq B\,n^{\alpha_s-2}\,\mu^n$.

The exponents $\gamma$ and $\alpha_s$ have been estimated
in the literature.  For example, the exponent $\gamma$ can be
accurately obtained from field theoretic calculations in three
dimensions.  Such calculations gave the very accurate estimate
\begin{equation}
\gamma = 1.15698 \pm 0.00034, \;\hbox{in reference \cite{LGZJ89}}.
\label{eqngamma}   
\end{equation}
The entropic exponent of lattice polygons in three dimensions
was estimated to have value
\begin{equation}
\alpha_s = 0.237\pm 0.002, \;\hbox{in reference \cite{GZJ98}}.
\label{eqnalphas}   
\end{equation}

\subsection{The conformational entropy of lattice polygons}

Let $\o{p}_n(\vec{r})$ be the number of polygons rooted at
$\vec{0}$ passing through the point $\vec{r} = (x,y,z)$ in $\mathL^d$.
Then $\w{p}_n(\vec{r}) = \o{p}_n - \o{p}_n(\vec{r})$ is the number 
of rooted lattice polygons avoiding the lattice site $\vec{r}$.  

The conformational entropy of a polygon rooted at the origin is given by
$S_n = k_B \log \o{p}_n$, where $k_B$ is Boltzmann's constant. 
The entropy of lattice polygons avoiding $\vec{r}$ 
is $\w{S}_n(\vec{r}) = k_B \log \w{p}_n(\vec{r})$.   Thus, the change in
entropy if the site $\vec{r}$ is excluded (or if a hard test particle 
is placed at $\vec{r}$) is given by
\begin{equation}
\Delta\,S_n(\vec{r}) = k_B \log \w{p}_n(\vec{r}) - k_B \log \o{p}_n,
\label{eqnS}    
\end{equation}
The change in extensive free energy is given by
$\Delta F_n(T) = - T\,\Delta\,S_n(\vec{r})$.   Thus, the entropic pressure
at the site $\vec{r}$ can be computed by considering the 
change in $F_n(T)$ due to a change in volume (by excluding the 
lattice site at $\vec{r}$).  This is given by 
\begin{equation}
P_n(\vec{r}) = \frac{\Delta F_n (T)}{\Delta V(\vec{r})}
=  -T \, \frac{\Delta\,S_n(\vec{r})}{\Delta V(\vec{r})} .
\end{equation}
Choose units so that $T=k_B=1$ and let $\Delta V(\vec{r})$
be the (unit square or cubical) volume element with $\vec{r}$ at 
its centre (so that $|\Delta V(\vec{r})| = 1$).  
By equation \Ref{eqnS} it follows that the pressure 
at the point $\vec{r}$ is given by
\begin{equation}
P_n(\vec{r}) = \log \o{p}_n - \log \w{p}_n(\vec{r})
= - \log \L  1 - \frac{\o{p}_n (\vec{r})}{\o{p}_n}\R .
\label{eqnP}    
\end{equation}
$P_n(\vec{r})$ is non-negative.  

Rescale $P_n(\vec{r})$ as follows.  Let $R^2_n = \LA r^2\RA_n$ be the  
mean square radius of gyration of rooted lattice polygons of length $n$.
Define the rescaled vectors $\vec{c}_r = \vec{r}/R_n$.  Then 
$a\,\vec{c}_r$ is a vector parallel to $\vec{r}$ for $a>0$.  
If $a=1$ then $\|\vec{r}\|_2 = R_n$, so that the units of $a$ is 
$R_n \simeq C n^\nu$.

Put $\mathbf{P}_n(a) = P_n(a\,\vec{c}_r R_n)$. 
Then $\mathbf{P}(1)$ is the pressure at a point a distance
$R_n$ from $\vec{0}$ in the direction $\vec{c}_r$.  Assuming that
$P_n(\vec{r})$ is isotropic shows that $\mathbf{P}_n(a)$ is the
pressure at a distance $aR_n$ from $\vec{0}$.  There are corrections
to this assumption for short polygons, but the data will show that this
is valid for longer polygons, subject to small corrections 
due to the geometry of the underlying lattice.  This defines the 
rescaled pressure $\mathbf{P}_n(a)$ in section \ref{section1},
and in equations \Ref{eqn1} and \Ref{eqn6}.

\subsection{Metric properties of walks and polygons}
\label{section21}   

The number of self-avoiding walks of length $n$ steps from 
$\vec{0}$ in $\mathL^3$ to a vertex $\vec{r}$ is $c_n(\vec{r})$.
Define $c_n(r) = \sum_{|\vec{r}|=r} c_n(\vec{r})$ to be the number 
of walks which ends at a distance $r$ from $\vec{0}$ after $n$
steps.  Since such walks end on a spherical shell of radius $r$,
it is expected that
\begin{equation}
c_n(r)  \simeq A_0\, r^2\, c_n(\vec{r}) 
\label{eqnC3}   
\end{equation}
in the cubic lattice $\mathL^3$.

The average distance of the endpoint of a self-avoiding walk of
length $n$ from the origin is denoted by $\LA r \RA_n$, and this
introduces a metric in the model.
The expected distance of the endpoint of the walk from
$\vec{0}$ can be computed from $c_n(r)$ by
\begin{equation}
R_c = \LA r \RA_n = \frac{1}{c_n} \sum_{r\geq 0} r\, c_n(r) \simeq C n^\nu
\label{eqnC4}   
\end{equation}  
where $\nu$ is the \textit{metric exponent} of the self-avoiding walk.
The metric exponent is known to high accuracy in three dimensions:
\begin{equation}
\nu = 0.587597\pm 0.000007\;\hbox{in reference \cite{C10}}
\label{eqnnu}   
\end{equation}
 
The ratio $c_n(\vec{r})/c_n$ is the (spherically symmetric)
end-to-end distribution function $P_n(r)$ of
self-avoiding walks \cite{F66}, and it is the probability that a walk of
length $n$ has endpoint a distance $r=|\vec{r}|$ from the origin
in the direction of $\vec{r}$.   $P_n(r)$ has behaviour given by
\begin{equation}
P_n(r) \sim R_c^{-d}\, F_0\L \sfrac{r}{R_c}\R
\end{equation}
That is, it is the inverse volume occupied by the end point of the 
walk (namely $R_c^{-d}$) multiplied by a modifying scaling function
$F_0$ which is a function of the distance from the origin scaled by 
(the length scale) $R_c$.  $F_0$ has the following 
conjectured properties \cite{dC74}:  
\begin{equation}
F_0(x) \sim x^g,\;\hbox{for small $x$ and}\;
F_0(x) \sim e^{-x^\delta} \; \hbox{for large $x$}.
\label{eqnC94}   
\end{equation}
where $g = \sfrac{\gamma-1}{\nu}$ \cite{dC74,dC80A} and
$\delta = \sfrac{1}{1-\nu}$ \cite{F66}.

Thus, in terms of $P_n(r)$,
\begin{equation}
c_n(\vec{r}) \simeq P_n(r)\,c_n \sim  \sfrac{1}{R_c^d}\, F_0\L \sfrac{r}{R_c}\R\, c_n
=  r^{-d} \L \sfrac{r}{R_c}\R^d \, F_0\L \sfrac{r}{R_c}\R\, c_n .
\end{equation}
Put $G(x) = x^dF_0(x)$, then the above becomes
\begin{equation}
c_n(\vec{r}) \approx C_0 \, r^{-d} G\L\sfrac{r}{R_c}\R\, c_n.
\label{eqncnvecr}   
\end{equation}
Notice that $R_c \sim R_n\simeq Cn^{\nu}$, where $R_n^2$ is the mean square
radius of gyration of lattice polygons.

\subsection{The number of polygons passing through a lattice point $\vec{r}$}

The pressure at a point $\vec{r}$ near a lattice polygon rooted at
$\vec{0}$ in $\mathL^d$ with $d\geq 3$
is computed by estimating the number of polygons passing
through $\vec{r}$.  The basic approach is to consider each such polygon 
as being composed of two walks from $\vec{0}$ to $\vec{r}$.
The number of such pairs of walks will be estimated by 
using equation \Ref{eqncnvecr}.

Denote the number of lattice polygons rooted at $\vec{0}$ in 
$\mathL^d$, of length $n$, containing the lattice point $\vec{r}$ 
by $\o{p}_n(\vec{r})$.   Then $\o{p}_n(\vec{r})$ can be approximated by
a pair of walks from $\vec{0}$ and ending in $\vec{r}$.  If these
walks avoid one another, then their union is a rooted lattice
polygon containing $\vec{r}$.  If they do not avoid one another, 
then an upper bound on $\o{p}_n(\vec{r})$ is obtained.  

This shows that
\begin{equation}
 \o{p}_n(\vec{r}) \simeq \sum_{k=0}^n P_{k,n-k}\,c_k(\vec{r})\, c_{n-k}(\vec{r}) .
\label{eqn14B}   
\end{equation}
where $P_{k,n-k}$ is the probability that a walk of length
$k$ and a walk of length $n{-}k$ from $\vec{0}$ to $\vec{r}$
avoid one another.

The summand in equation \Ref{eqn14B} can be estimated by using
equation \Ref{eqncnvecr}.  This replaces the functions
$c_k(\vec{r})$ and $c_{n-k}(\vec{r})$ in the summand of
equation \Ref{eqn14B} by factors containing $c_k$ and $c_{n-k}$.

The probability $P_{k,n-k}$ is estimated as follows. Let 
$Q_{k,n-k}(\vec{r},\vec{s})$ be the probability that a walk of 
length $k$ from $\vec{0}$ to $\vec{r}$ intersects a walk of length
$n{-}k$ from $\vec{0}$ to $\vec{s}$.  Since such walks are likely
to intersect near $\vec{0}$ (where they start, irrespective of the 
choices for $\vec{r}$ and $\vec{s}$), this probability is 
to leading order given by the probability that two walks, 
of lengths $k$ and $n{-}k$, intersect each other.  That is, to
leading order $Q_{k,n-k}(\vec{r},\vec{s})$ should be only weakly
dependent on $\vec{r}$ and $\vec{s}$.   In other words,
$P_{k,n-k}$ may be approximated by the probability that two walks
from $\vec{0}$, of lengths $k$ and $n{-}k$ avoid one another.
Denote this probability by $P_{k,n-k}^r$.  This gives the following estimate for $\o{p}_n(\vec{r})$:
\begin{equation}
 \o{p}_n(\vec{r}) \simeq 
C_0^2 r^{-2d}
\sum_{k=0}^nP^r_{k,n-k}\, G(r/Ck^\nu)\,
G(r/C(n{-}k)^\nu)\, c_k\,c_{n-k} .
\label{eqn14BCA}   
\end{equation}
$P^r_{k,n-k}$ is given by $P^r_{k,n-k} = c_{n}/(c_k\,c_{n-k})$.  Thus, 
substituting $P^r_{k,n-k}$,
\begin{equation}
 \o{p}_n(\vec{r}) \simeq A_1\, c_n\, r^{-2d}
\sum_{k=0}^n  G(r/Ck^\nu)\,
G(r/C(n{-}k)^\nu)\, 
\label{eqn14BC}   
\end{equation}
for some constant $A_1 = C_0^2$.

Introduce the scaling factor $a$ by putting 
$r=a\,R_n$ where $R_n\simeq C n^\nu$ is the root of the mean square 
radius of gyration of the polygon.  Then 
$\vec{r} = \L a\,R_n \R\, \sfrac{\vec{r}}{|\vec{r}|}$ and $\o{p}_n(\vec{r})$
becomes a function of $a$.  Put
\begin{equation}
\widehat{p}_n(a) =  \o{p}_n(\L a\,R_n \R \sfrac{\vec{r}}{|\vec{r}|}).
\end{equation}  
Then $\widehat{p}_n(a)$ is the number of polygons passing through a point
a distance $a\,R_n$ from the origin, where $R_n\simeq C n^\nu$ defines
the length scale in the problem.  The result is that
\begin{equation} 
G(r/Ck^\nu) \approx G(a\, (\sfrac{k}{n})^{-\nu})\;
\hbox{and}\,
G(r/C(n{-}k)^\nu) \approx G(a\, (1{-}\sfrac{k}{n})^{-\nu})
\end{equation}
in equation \Ref{eqn14BC}.

Substitute $c_n$ by its scaling relation in equation \Ref{eqnC1},
and put $A_2 = A A_1C^{-2d}$
to see that 
\begin{equation*}
 \widehat{p}_n(a) \simeq  A_2\,n^{\gamma-1} 
\L a\, n^\nu \R^{-2d} \mu^n
\sum_k G(a\, (\sfrac{k}{n})^{-\nu})\,
G(a\, (1{-}\sfrac{k}{n})^{-\nu}).
\end{equation*}
Approximate the summation by an integral over $k$, and 
change variable $x =\sfrac{k}{n}$.  This gives
\begin{equation*}
 \widehat{p}_n(a) \simeq  A_2 
\, n^{\gamma-2d\nu}
a^{-2d} \,\mu^n
\int_0^1
G(a/x^\nu)\,
G(a/(1{-}x)^\nu) \,dx .
\end{equation*}

\begin{figure}[t!]
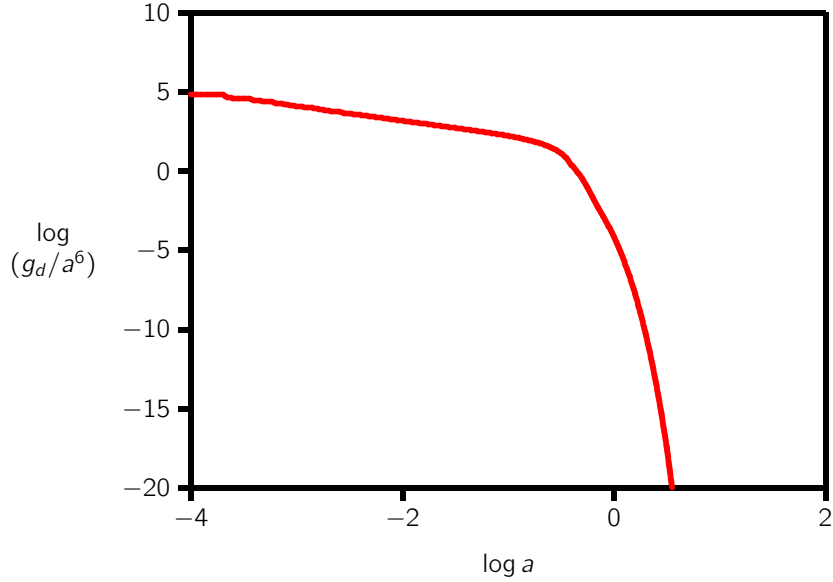

\input figureA.tex
\caption{A plot of $\log\L g_d(a)\,a^{-6} \R$ against
$\log a$.}
\label{figureA}    
\end{figure}

Define the function
\begin{equation}
g_d(a) = \int_0^1 
G(a/x^\nu)\,G(a/(1{-}x)^\nu) \,dx .
\label{eqnga}   
\end{equation}
where $G(x)$ is given by $G(x) = x^dF_0(x)$ and 
$F_0(x)$ is defined in equation \Ref{eqnC94}. Then 
\begin{equation}
\widehat{p}_n(a) \simeq A_2\,  \mu^n\, n^{\gamma-2d\nu} \, g_d(a)/a^{2d}
\label{eqnB3}   
\end{equation}
for some constant $A$. 
The dependence of $\widehat{p}_n(a)$ on $a$ is given by $g_d(a)/a^{2d}$.
This decays quickly to zero with increasing $a$.  In figure \ref{figureA}
a plot of $g_d(a)/a^{2d}$ against $a$ is shown on a logarithmic axes
in $d=3$ dimensions.  For small
values of $a$ the graph decreases slowly on the log scale, but it turns over 
at about $a=1$ and then quickly decays to zero.  This happens because of the
super-exponential decay of the function $G(x)$ (see equation \Ref{eqnC94}).

\subsection{The rescaled pressure $\mathbf{P}_n(a)$}

By equation \Ref{eqnP} the pressure due to rooted polygons at a point
a distance $a\,R_n$ from the origin is approximately given by
\begin{equation}
\mathbf{P}_n(a) \simeq - \log \L 1- \frac{\widehat{p}_n(a)}{\o{p}_n}\R,
\label{eqn29}   
\end{equation}
where $\widehat{p}_n(a)$ is the number of polygons passing 
through a point a distance $aR_n$ from the origin
and $\o{p}_n = n\,p_n$ is the number of rooted polygons
of length $n$.

The scaling of the pressure is obtained by using equations 
\Ref{eqnB3} and \Ref{eqnP1}.  In particular, for $a$ not too small
and for $n$ large this becomes
\begin{equation}
\mathbf{P}_n(a)  \simeq - \log \L 1 - \frac{A_2\, g_d(a)}{B\, a^{2d}}
\, n^{\gamma-2d\nu-\alpha_s + 2} \R ,
\end{equation}
since $\o{p}_n = n\,p_n$.
Since $\gamma{-}2d\nu{-}\alpha_s{+}2 < 0$, and since $a^{-2d}g_d(a)$
quickly approaches zero, the logarithm may be expanded.  To
leading order
\begin{equation}
\mathbf{P}_n(a) \simeq \frac{A_2\, g_d(a)}{B\, a^{2d}}
\, n^{\gamma-2d\nu-\alpha_s + 2} .
\label{eqnPscaleG}    
\end{equation}
The hyperscaling relation $2{-}\alpha_s = d\nu$ for polygons
in three and four dimensions can be used to obtain
\begin{equation}
\mathbf{P}_n(a) \simeq \frac{A_2\, g_d(a)}{B\, a^{2d}}
\, n^{\gamma-d\nu} .
\label{eqnPscaleGH}    
\end{equation}
Using the approximate values $\gamma\approx 1.15$
and the Flory value $\nu=0.6$ gives 
$\mathbf{P}_n(a) \sim n^{-0.65}$, while
the more accurate estimates in equation \Ref{eqngamma} and 
\Ref{eqnnu} gives $\mathbf{P}_n(a) \sim n^{-0.606}$.
This gives the following relation in $d=3$ dimensions:
\begin{equation}
n^\rho\, \mathbf{P}_n(a) \simeq C\,g_d(a)a^{-6} ,
\label{eqnPscale}    
\end{equation}
for some constant $C$, where the exponent $\rho$ should be
approximately equal to
$0.606$ (and it is equal to $\gamma {-} d\nu$ in $d\geq 3$ dimensions
-- in contrast with its value in $2$ dimensions given in equation
\Ref{eqn3}).

This prediction
may be tested numerically by rescaling the lattice by $n^{-\nu}$
and the pressure by $n^\rho$ and then plotting
$[n^\rho\, \mathbf{P}_n(a)]$ against $[|\vec{r}|/n^\nu]$.  This
should collapse data for a range of choices of $\vec{r}$ and $n$
to a single curve which is only a function of $a$ (and has the
general shape shown in figure \ref{figureA}).

{\renewcommand{\baselinestretch}{1.5}
\begin{table}[t!]
\begin{center}
\caption{Numerical estimates of $\o{p}_n$ and $\w{p}_n(1,0,0)$ for $3_1$}
\label{Tabledata}   

\begin{tabular}{||r|ll|ll||}
\hline
$n$\q &\q $\o{p}_n$ &\q $\sigma_n$ &\q $\w{p}_n(1,0,0)$ &\q $\sigma_n$ \\
\hline\hline
$24$      &$33936$   &$0$ &$16797$ &$21$ \\
$26$      &$3.1067\times10^6$   &$0.0094\times 10^6$ &$1.5230\times10^6$ &$0.0182\times 10^6$ \\
$28$      &$1.7139\times10^8$   &$0.0091\times 10^8$ &$8.4007\times10^7$ &$0.0102\times 10^7$ \\
$30$      &$7.3863\times10^9$   &$0.0052\times 10^9$ &$3.6367\times10^9$ &$0.0048\times 10^9$ \\
$32$      &$2.7385\times10^{11}$   &$0.0024\times 10^{11}$ &$1.3577\times10^{11}$ &$0.0018\times 10^{11}$ \\
$34$      &$9.1671\times10^{12}$   &$0.0087\times 10^{12}$ &$4.5805\times10^{12}$ &$0.0062\times 10^{12}$ \\
\hline
\end{tabular}

\end{center}
\end{table}
}

\section{The pressure near lattice knots}
\label{section3}

Lattice polygons in $\mathL^3$ are 
piecewise linear embeddings of the circle 
in three space, and have well-defined knot type.  For example,
all polygons of length $4$ are unknotted, and have
knot type the \textit{unknot} (denoted by $0_1$).
It is known that all lattice polygons in $\mathL^3$ of length
$n<24$ are unknotted \cite{D93}.

Polygons in $\mathL^3$ of length $24$ or longer may have 
non-trivial knot type.  For example, polygons of knot type
the \textit{trefoil} (denoted by $3_1$) can be realised at length $n=24$
in $\mathL^3$ \cite{D93}, and moreover, there are precisely
$3328$ distinct embeddings of length $24$ which
have knot type $3_1$ \cite{SIA09}.    The \textit{minimal length} of polygons
of knot type $3_1$ in $\mathL^3$ is $n_{min}(3_1) = 24$.
(Similarly, $n_{min}(0_1) = 4$ for the unknot).  The knot type of
the schematic ring polymer in figure \ref{figure1} is a trefoil.
If a polygon does not have knot type
$3_1$ and its length is less than $30$, then its knot type
is $0_1$ \cite{SIA09}. 

Lattice knots of type $4_1$ (the \textit{figure eight knot}) can
be realised with minimal length $n_{min}(4_1) = 30$, of
type $5_1$ at $n_{min}(5_1) = 34$ and $5_2$ at
$n_{min}(5_2) = 36$  -- see for example references
\cite{SIA09,JvRR11B,JvRR12} for more results.

\begin{figure}[t!]
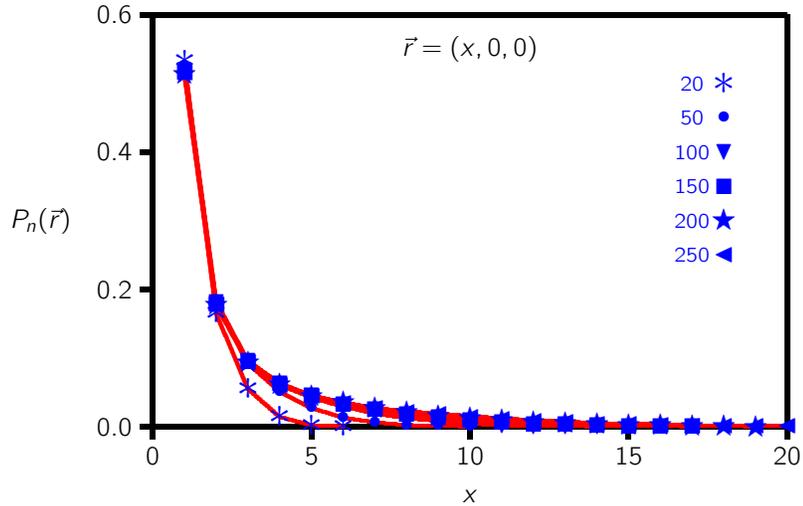

\input figureB.tex
\caption{The pressure $P_n(x,0,0)$ (equation \Ref{eqnP})
along the $X$-axis at the lattice sites $(x,0,0)$  for $x\in\{1,2,\ldots,30\}$,
and for $n=20$(\textasteriskcentered), $n=50$($\bullet$),
$n=100$($\blacktriangledown$), $n=150$($\blacksquare$),
$n=200$($\bigstar$) and $n=250$($\blacktriangleleft$).
}
\label{figureB}    
\end{figure}

The rescaled pressure $\mathbf{P}_n(a)$ for lattice polygons
in equation \Ref{eqnPscale} also applies to 
knotted polygons.  To see this, notice that $\o{p}_n(K)$, 
the number of lattice polygons of knot type $K$, rooted 
at $\vec{0}$, has been shown to have asymptotic growth given by
\begin{equation}
\o{p}_n(K) \approx B_k n^{\alpha_s - 2 + N_K} \mu^n
\label{eqn34}  
\end{equation}
where $N_K$ is the number of prime components in $K$.
In the case of prime knot types, $N_K=1$, and for the unknot
$N_K=0$ (see reference \cite{MOSZ04,OSV09,OTJW96,OTJW98}).

This may be understood as follows:  The average structure
of a knotted polygon in the scaling limit is that of an unknotted 
polygon with small knotted ball-pairs distributed along its length.
Each knotted ball-pair is an arc of prime knot type with endpoints on
a topological ball containing it, and
this gives a factor of $n$ for each such ball-pair, giving rise
to a factor $n^{N_K}$ in equation \Ref{eqn34} -- see references
\cite{OTJW96,OTJW98} for more details.  

Thus, for a polygon of knot type $K$ equation \Ref{eqn14B}
is modified to 
\begin{equation}
 \o{p}_n(\vec{r},K) \simeq  C_K n^{N_K} 
\sum_{k=0}^n P_{k,n-k}\, c_k(\vec{r})\, c_{n-k}(\vec{r})  .
\label{eqn14CC}   
\end{equation}
where $\o{p}_n(\vec{r},K)$ is the number of polygons of length
$n$, rooted at $\vec{0}$, of knot type $K$ with $N_K$
prime knot factors, and passing through the vertex $\vec{r}$.
The factor of $n^{N_K}$ is carried through without modification
to equation \Ref{eqnB3}.   In the case of unknotted polygons
$N_K=0$, and the pressure has the scaling form given in 
equation \Ref{eqnPscale}.  However, if $N_K \not=0$, then
the pressure near a knotted polygon is computed from
equation  \Ref{eqn29}.  Notice that the factor $n^{N_K}$ cancels
in the numerator and demoninator (which is given by equation 
\Ref{eqn34}).  Thus, the scaling of the pressure is unchanged,
and for knotted polygons it is given by equation \Ref{eqnPscale}.

The GAS algorithm \cite{JvRR09} was used to approximately enumerate
rooted knotted lattice polygons avoiding fixed vertices $\vec{r}$
in the cubic lattice.  The algorithm was tuned to sample polygons of
lengths up to $n=400$, sampling along $500$ independent sequences,
each of length $6\times 10^7$ attempted BFACF moves.  This gives
a total of $3 \times 10^{10}$ iterations for each knot type in
the set $\{0_1,3_1,4_1,5_1,5_2\}$ (where $0_1$ is the
unknot, $3_1$ is the trefoil, and $4_1$ is the figure eight knot).  

A sample of data for the trefoil knot is shown in table \ref{Tabledata}.
Data were collected on points in the $xy$-plane of $\mathL^3$
with Cartesian coordinates of the form $(x,y,0)$ with
$0\leq x \leq 29$ and $1\leq y \leq 30$.   By symmetry this gives
estimated pressures in the $xy$-plane for $-30\leq x \leq 30$ and
$-30 \leq y \leq 30$.

\begin{figure}[t!]
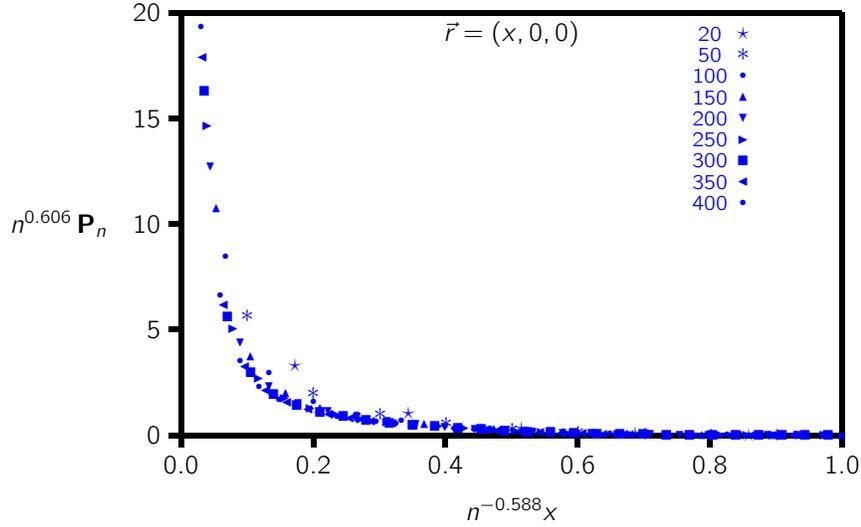

\input figureC.tex
\caption{Testing the scaling prediction in equation 
\Ref{eqnPscale}.  The rescaled pressure $n^{0.606}P_n(x,0,0)$ is
plotted as a function of $n^{-0.588} \|\vec{r}\|_2$.  These data include 
all the data points in figure \ref{figureA}.  Differently shaped
points correspond to different values of $n$, as shown in the key.
The data collapse to a single curve, uncovering the scaling 
function $g_d(a)/a^6$ in equation \Ref{eqnPscale}.}
\label{figureC}    
\end{figure}

\begin{figure}[h!]
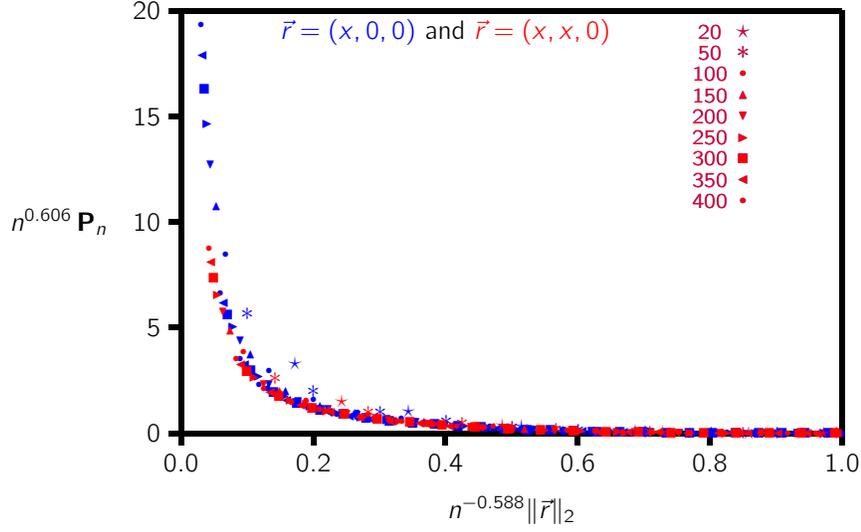

\input figureCxx.tex
\caption{The rescaled pressure along points $\vec{r}=(x,0,0)$
and $\vec{r}=(x,x,0)$ for $x\in\{1,2,\ldots,30\}$. 
The data of figure \ref{figureC} are represented 
as blue points.  Red data points represent pressures along the
diagonal $\vec{r}=(x,x,0)$.  The data collapse to a single curve, so 
that the pressure decreases and scales at the same rate along 
these two directions.  This is support for the suggestion in 
figure \ref{figure2} that the pressure field is isotropic.}
\label{figureCxx}    
\end{figure}

\subsection{The pressure of the unknot}

The pressures for the unknot were computed from data for unknotted
polygons similar to the data in table \ref{Tabledata} by using
equation \Ref{eqnP}.  In figure \ref{figureB} the pressure of
unknotted polygons along points on the $x$-axis is plotted as
a function of distance from the origin -- that is, $P_n(x,0,0)$ is 
plotted as a function of $x$ for $n\in\{20,50,100,150,200,250\}$.

The data in figure \ref{figureB} show that the pressure increases quickly on 
approaching the polygon, which is rooted at the origin $\vec{0}$ in
$\mathL^3$.  Since the origin is always occupied by the polygon, the 
pressure here is infinite (this is an artifact of the model).  

Rescaling of the data in figure \ref{figureB} according to 
equation \Ref{eqnPscale} shows that $n^\rho\,\mathbf{P}_n(a)$
should be plotted against $n^{-\nu}\|\vec{r}\|_2$.  The scaling
analysis of section \ref{section2} predicts that this will
collapse the data in figure \ref{figureB} to a single curve.

Using the numerical estimates for $\gamma$ in equation
\Ref{eqngamma} and for $\nu$ in equation \Ref{eqnnu}
gives $\rho \approx 0.606$ (and $\nu \approx 0.588$).
Thus, the rescaled data $n^{0.606}\mathbf{P}_n$ should be
plotted against $n^{-0.588}\| \vec{r} \|_2$.  This is done in
figure \ref{figureC}, which includes all the data in figure \ref{figureB}
as well as additional data from unknotted polygons of length
$n\in\{300,350,400\}$.  These results show that the data rescale
to a single curve with increasing values of $n$ -- this is strong
evidence in support of the scaling analysis in section \ref{section2}.

\begin{figure}[t!]
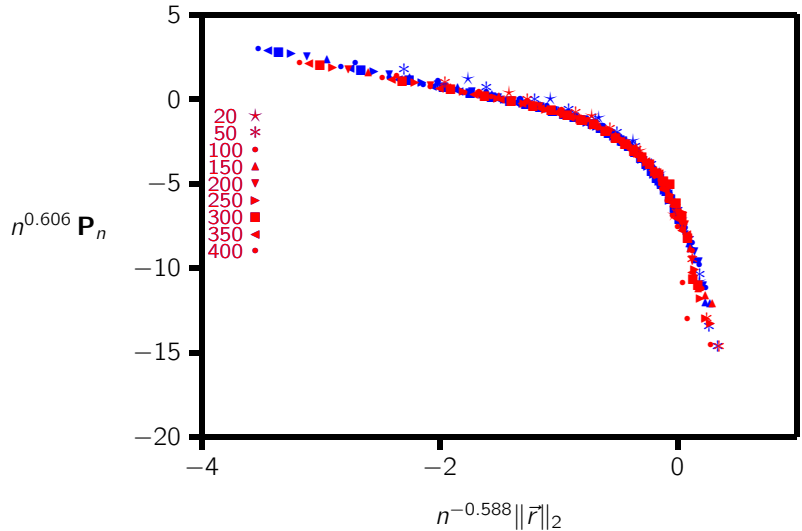

\input figureFlogx0xx.tex
\caption{The data of figure \ref{figureCxx} on a log-log
scale.  The rescaled pressure along points $\vec{r}=(x,0,0)$ are in blue,
and along $\vec{r}=(x,x,0)$ are in red, for $x\in\{1,2,\ldots,30\}$. 
The data collapse to a single curve, so 
that the pressure decreases with distance from the origin
in the same way along these directions.}
\label{figureFlogx0xx}    
\end{figure}

The plot of the pressure field near an unknotted polygon of length
$n=100$ in figure \ref{figure2} strongly suggests that the field is
isotropic.  This can be tested numerically by examining the rescaled
pressure along other directions in the lattice.  In particular, data
were collected at points $\vec{r}=(x,x,0)$ along a diagonal
in $\mathL^3$. The resulting plot is illustrated in figure \ref{figureCxx}:
The data along the two directions collapse to a single curve in
the rescaled coordinates, supporting the suggestion that the 
pressure field is isotropic, even at small distances in $\mathL^3$.

The scaling function $g_d(a)/a^6$ in equation \Ref{eqnPscale} 
is uncovered in figures \ref{figureC} and \ref{figureCxx}.  As expected,
it decays very quickly with increasing (rescaled) distance.  The expected
shape of this scaling function is illustrated in figure \ref{figureA}
on logarithmic axes.  
In figure \ref{figureFlogx0xx} the data of figure \ref{figureCxx} is plotted
on log-log axes.  The shape of the rescaled data is similar to that
of the theoretically derived shape of $g(a)/a^6$ shown in figure
\ref{figureA}.  As expected, the data present a picture of relative high 
entropic pressure near the lattice polygon, but this crosses through a 
boundary layer at distance roughly 
$r \approx n^\nu$ (or $a\approx 1$) to a phase
where the pressure decreases quickly with increasing distance
-- the crossover happens at the ``boundary" of the polygon, which 
may be considered a gas of monomers localised inside its mean radius.

The turnover at the ``boundary" of the polygon
defines a ``surface layer".  At distances inside the surface layer
there is a non-zero density of vertices occupied by the lattice knot,
giving rise to non-zero entropic pressure which decreases
at an exponential rate with increasing distance.
Once the surface layer is crossed the rate of decrease in the entropic
pressure increases dramatically, and it falls off to zero quickly with 
increasing distance (as seen in the steep descend of $n^\rho \mathbf{P}_n$ 
with increasing distance in figure \ref{figureFlogx0xx} for $a>1$).

\begin{figure}[t!]
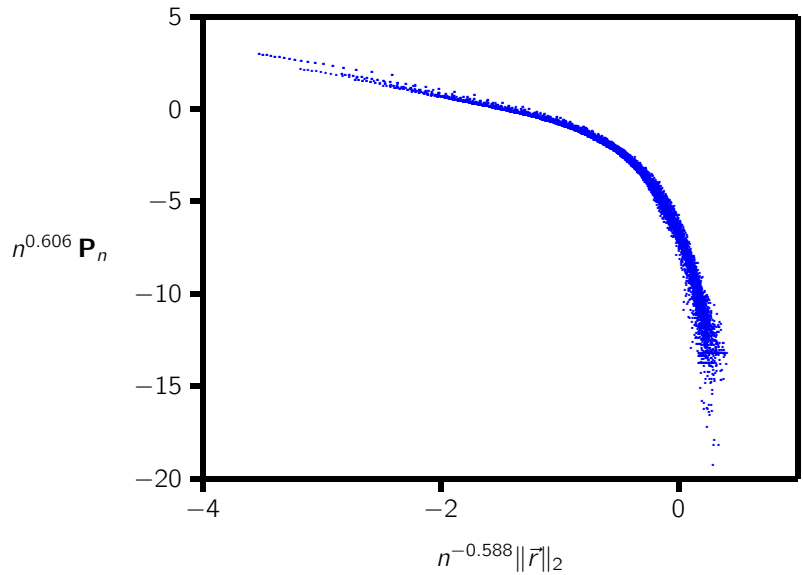

\input figureFlog-all.tex
\caption{The rescaled pressure $n^{0.606}P_n(x,0,0)$ 
plotted as a function of $n^{-0.588} x$ on a log-log scale.  This scatter
plot is for all polygons of lengths $4n$ with $100\leq 4n\leq 400$
and for lattice points $(x,y,0)$ (where $1\leq x \leq 30$ 
and $0 \leq y \leq 29$).  The data uncover the scaling function
$\log(g_d(a)/a^6)$ (see equation \Ref{eqnPscale}).}
\label{figureFlog-all}    
\end{figure}

\begin{figure}[h!]
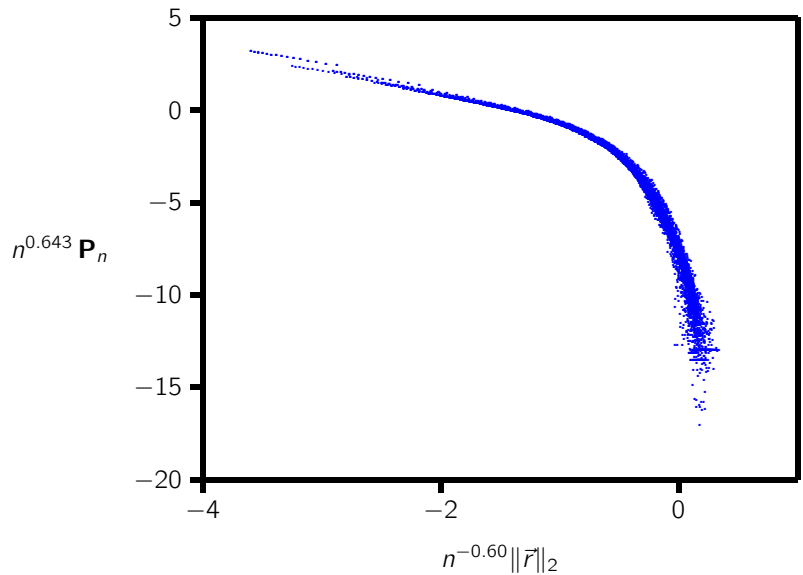

\input figureFF-Flory.tex
\caption{A plot similar to figure \ref{figureFlog-all}, but using the
Flory value $\nu = 0.60$ instead.}
\label{figureFF-Flory}    
\end{figure}

A particle approaching the lattice knot from a large distance experiences
little pressure at first.  The pressure increases steeply as the surface layer
of the lattice knot is approach.  Once the surface layer is crossed the
relative pressure gradient reduces to the more gentle slop seen for 
small values of $a$ in figure \ref{figureFlogx0xx} -- but the increase
is still at an exponential rate.    The gradient of the pressure 
induces a repulsive entropic force on an approaching particle --
a particle near the lattice knot will be expelled from its vicinity 
by these induced forces.

In figure \ref{figureFlog-all} all the rescaled data for polygons of lengths
$100 \leq 4n \leq 400$ are plotted on logarithmic axes, for all lattice points
$(x,y,0)$ (where $1\leq x \leq 30$ and $0 \leq y \leq 29$).  The data
accumulate in the vicinity of a single curve of shape given by
figure \ref{figureA}. 

A second graph is presented in figure \ref{figureFF-Flory}
where the Flory value for $\nu$ (this is $\nu=0.60$) was used, 
and with $\gamma=1.15$. Thus, the rescaled pressure 
$n^{0.643}\mathbf{P}_n(a)$ is plotted against $n^{-0.60} \| x\|_2$.  
The result is a plot very similar to figure \ref{figureFlog-all}, showing 
that the scaling is also consistent with the Flory value of $\nu$.

\begin{figure}[t!]
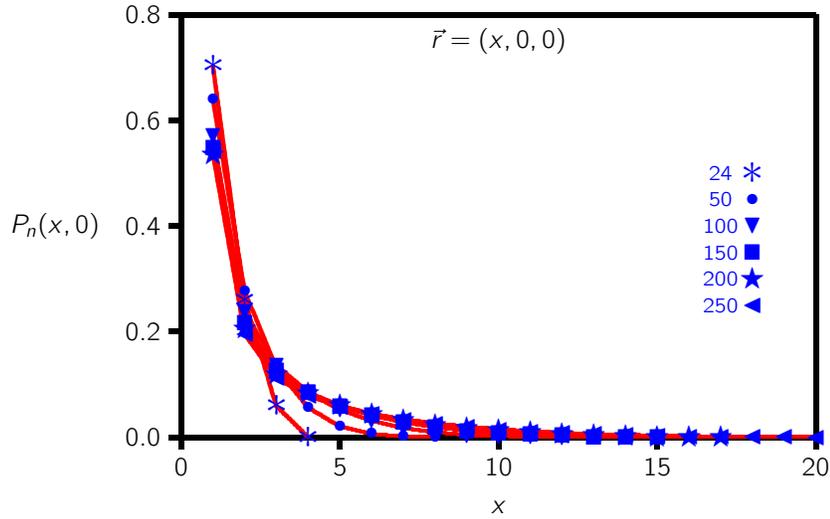

\input figureB31.tex
\caption{The pressure $P_n(x,0,0)$ of lattice trefoils
(equation \Ref{eqnP})
along the $X$-axis at the points $(x,0,0)$  for $x=1,2,3,\ldots,30$,
and for $n=24$(\textasteriskcentered), $n=50$($\bullet$),
$n=100$($\blacktriangle$), $n=150$($\blacksquare$) and
$n=200$($\bigstar$) and $n=250$ ($\blacktriangleleft$).
}
\label{figureB31}    
\end{figure}

\begin{figure}[h!]
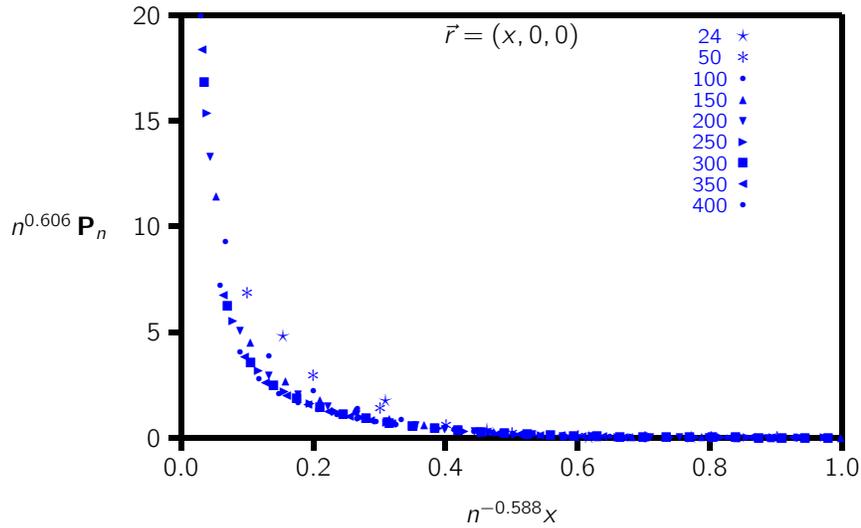

\input figureC31.tex
\caption{Testing the scaling prediction in equation 
\Ref{eqnPscale} for lattice trefoils.  The rescaling of
data in figure \ref{figureB31} is done by plotting $n^{0.606}P_n(x,0,0)$ is
plotted as a function of $n^{-0.588} x$.  The data collapse to
a single curve, uncovering the scaling function $g_d(a)/a^6$
in equation \Ref{eqnPscale}.  This plot contains all the data of figure
\ref{figureB31} as well as data at larger values of $n$.}
\label{figureC31}    
\end{figure}

\begin{figure}[t!]
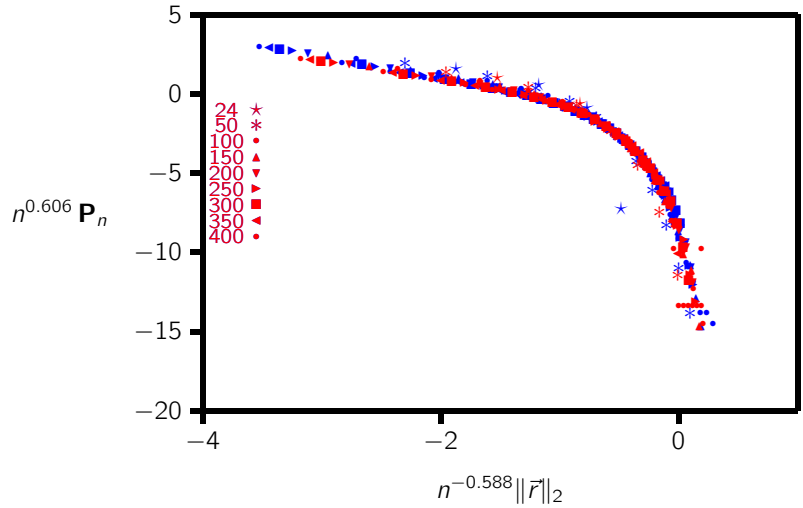

\input figureFlog31x0xx.tex
\caption{ The rescaled pressure $n^{0.606}P_n$ 
plotted as a function of $n^{-0.588} \|\vec{r}\|_2$ on a log-log scale
for lattice trefoil knots.  The data include the pressures at vertices
with coordinates $(x,0,0)$ and $(x,x,0)$ for $x\in\{1,2,\ldots,30\}$. 
This graph also includes all the data in figure \ref{figureC31}. 
 The data collapse to a single curve, uncovering $\log(g_d(a)/a^6)$
in equation \Ref{eqnPscale}.}
\label{figureFlog31x0xx}   
\end{figure}

\begin{figure}[bh!]
\input figureFlog31-01.tex
\caption{ The rescaled pressure $n^{0.606}P_n(x,0,0)$ 
plotted as a function of $n^{-0.588} \|\vec{r}\|_2$ on a log-log scale
for the unknot (blue) and the trefoil (red).  At small distances the
the data coincide, but the points separate for $a$ approaching
one, with pressure dropping more quickly for the trefoil than for the
unknot. }
\label{figureFlog31-01}    
\end{figure}

\begin{figure}[th!]
\centering
\includegraphics[width=0.55\textwidth,trim=0 10 25 25]{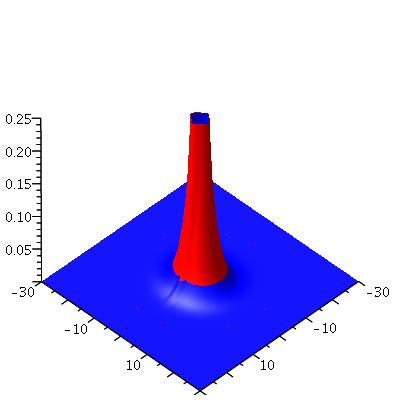}
\caption{The pressure field near a lattice knots of type unknot and type
trefoil plotted on the same graph.  Data for the unknot are shown in blue,
and for the trefoil knot in red.  This shows that the pressure of the trefoil
knot is higher close to the origin (at short distances from the knot --
this is the red data), and lower than the pressure of the unknot 
at larger distance (where the blue data is above the red data).  This is consistent
with the trefoil being a more compact knot, exploring states with vertices
closer to the origin (and increasing the pressure there), while not visiting
vertices at larger distances at the same rate than the unknotted polygons
(therefore lowering the pressure at larger distances).}
\label{figure2T}    
\end{figure}

\subsection{The pressure near a lattice trefoil}

The entropic pressure near a lattice trefoil knot rooted at
the origin were computed at points $(x,0,0)$ for $1\leq x\leq 30$ in
the same way as data were collected for unknotted polygons. 
Some data are plotted in figure \ref{figureB31} -- note that the shortest
polygons have length $24$, which is the minimal length of a cubic lattice
trefoil \cite{D93}.

The data in figure \ref{figureB31} can be rescaled (similar to the
unknot data) as was done in figure \ref{figureC}.  The result is displayed in figure
\ref{figureC31}.  As for the unknot,  the pressure curves
in figure \ref{figureB31} collapse to a single curve, exposing the
scaling function for lattice trefoils.

The rescaled pressure of lattice trefoils were also plotted
on logarithmic axes in figure \ref{figureFlog31x0xx}.  This 
graph includes the pressure on the vertices $(x,0,0)$ and $(x,x,0)$ with
$1\leq x \leq 30$.   The result is a graph similar to the graph in figure 
\ref{figureFlogx0xx}, showing a pressure profile like that of the unknot,
with a well-defined surface layer.  At the surface layer the relative
pressure gradient is large, quickly decaying with distance from the 
origin -- this relative gradient is more gentle once the surface layer 
is crossed to vertices closer to the origin.

A comparison of the pressures of the unknot and of the trefoil knot
may be done by plotting figures \ref{figureFlogx0xx} and \ref{figureFlog31x0xx}
on the same axes.  This is done in figure \ref{figureFlog31-01}, with
the blue data points corresponding to the unknot, and the red data points
corresponding to the trefoil.  The data coincide at short distances, as
expected, but as $a$ approaches one the trefoil data falls away 
marginally more quickly, showing that the trefoil knot is slightly smaller
than the unknot, and is therefore less likely to occupy vertices at larger
distances from the origin. 

In figure \ref{figure2T} the pressures of the unknot and the trefoil
are compared for polygons of length $n=100$.  The blue data are identical to
the data in figure \ref{figure2} -- this is the pressure of the unknot 
interpolated at points $(n,m,0)\in\mathL^d$ for $-30\leq n,m\leq 30$.
The red data are (similarly) the interpolated pressures of polygons of knot type
$3_1$.  At short distances the red data is above the blue, and vice
versa at longer distances.  This shows that the entropic pressure of
a trefoil polygon is higher than that of the unknot at points close to the orgin.
At large distances the converse is seen, the unknot has a higher entropic
pressure than the trefoil.  This result is consistent with the unknot
being a more expanded knot, more likely to occupy vertices at relative
large distances from the origin, compared to the trefoil.

\subsection{The pressures of other knots}

Simulations were similarly performed to compute the entropic
pressure near lattice polygons of knot types $4_1$, $5_1$ and $5_2$.
These cases should have the same scaling properties as seen for
the trefoil (for example in figure \ref{figureFlog31x0xx}).

\begin{figure}[t!]
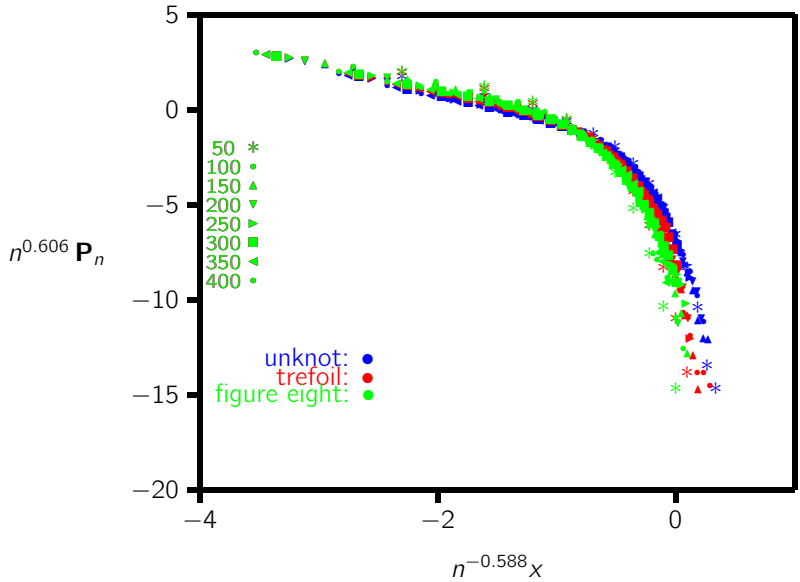

\input figureFlog31-41.tex
\caption{ The rescaled pressure $n^{0.606}P_n(x,0,0)$ 
plotted as a function of $n^{-0.588} \|\vec{r}\|_2$ on a log-log scale
for the unknot (blue), the trefoil (red) and the figure eight knot
(green).  At small distances the data scale similarly, with the largest
pressure seen for the figure eight knot.  At large distances the
pressure for the figure eight falls off more quickly than for the
trefoil, which in turn decreases more quickly than for the unknot.}
\label{figureFlog31-41}    
\end{figure}

In figure \ref{figureFlog31-41} the rescaled entropic pressures
of the unknot, the trefoil and the figure eight knot are plotted against
the rescaled length $n^{-\nu}\|\vec{r}\|$ for $\vec{r}=(x,0,0)$ on
a log-log scale.  The data for the unknot (blue) and the trefoil (red)
are the same as in figure \ref{figureFlog31-01}.  Data for the figure
eight knot are shown in green.  These data show that for small distances
the pressures are almost equal.  However,  at larger distances the
data for the figure eight decreases marginally more quickly than
for the trefoil.  That is, the figure eight knot has smaller pressure as the surface
layer is crossed, when compared to the trefoil knot (which in turn has smaller
pressure at the surface layer when compared to the unknot).

\begin{figure}[th!]
\input figureFlog01--52.tex
\caption{ The rescaled pressure $n^{0.606}P_n(x,0,0)$ 
plotted as a function of $n^{-0.588} \|\vec{r}\|_2$ on a log-log scale
for the knot types $5_1$ (cyan) and $5_2$ (orange).}
\label{figureFlog01--52}    
\end{figure}

Similar results are observed when data for $5_1$ and $5_2$ are
added into figure \ref{figureFlog31-41}.  This is shown in figure
\ref{figureFlog01--52} -- the five crossing knots are slightly more
compact than the figure eight knot and so there is a slight excess
pressure at small distances and a slight deficit of pressure at larger
distances in the surface layer.

The two five crossing knots have (up to the accuracy in this study), 
virtually the same pressure curve, as seen in figure \ref{figureFlog51-52},
but plotting the pressure for polygons of length $n=100$ and knot 
types $5_1$ and $5_2$ in figure \ref{cone5152} show that the pressure
of $5_2$ at short distances exceeds that of $5_1$.  At longer distances (near
the surface layer), $5_1$ has higher pressure.  The uneven boundary between 
the green and yellow regimes, and the patches of yellow, are due to noise 
in the data.

\begin{figure}[t!]
\input figureFlog51-52.tex
\caption{ The rescaled pressure $n^{0.606}P_n(x,0,0)$ 
plotted as a function of $n^{-0.588} \|\vec{r}\|_2$ on a log-log scale
for the knot types $5_1$ (cyan) and $5_2$ (orange).}
\label{figureFlog51-52}    
\end{figure}

\begin{figure}[th!]
\centering
\includegraphics[width=0.55\textwidth,trim=0 10 25 25]{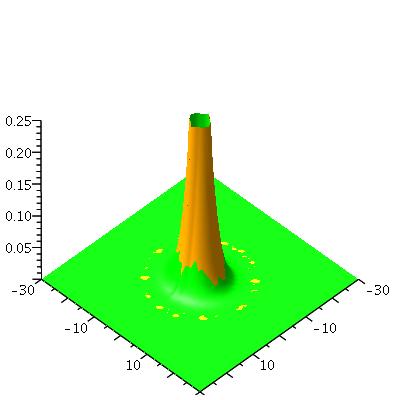}
\caption{The pressure cone for the knots $5_1$ (green) and $5_2$ (yellow)
near rooted polygons of length $n=100$.  
These data show that the pressure due to $5_2$ exceeds that of
$5_1$ at short distances, but that $5_1$ has larger pressure at longer
distances.  The uneven barrier between the colours, and the patches of
yellow are due to noise in the data.}
\label{cone5152}    
\end{figure}

\section{Conclusions}
\label{section4}

In this paper the entropic pressure near a self-avoiding walk model
of a ring polymer was modelled.  This study extends the results in
reference \cite{GJvR13} to the cubic lattice -- the scaling argument
in that reference is shown here not to be valid in three dimensions.
Instead, a slightly modified approach produced the scaling relation
in equation \Ref{eqnPscaleG} which may be written as 
$n^\rho \mathbf{P}_n(a) \simeq C g_d(a)/a^6$ where $a$ is 
length rescaled by $n^\nu$ and $\rho=\gamma-d\nu$ in 
$d\geq 3$ dimensions.

This scaling relation was tested by collecting data on unknotted
polygons rooted at the origin in $\mathL^3$.  The data 
gave results consistent with the scaling relation, as seen in 
figures \ref{figureC}, \ref{figureCxx}, \ref{figureFlogx0xx}
and \ref{figureFlog-all}.  In fact, the data in figures
\ref{figureFlogx0xx} and \ref{figureFlog-all} uncover the
predicted shape of the scaling function $g_d(a)/a^{6}$
plotted in figure \ref{figureA}.

Similar results were obtained for knotted polygons.  The expectation
is that the pressure of a knotted polygon should have the same
scaling behaviour as seen for the unknot, and figures
\ref{figureC31} and \ref{figureFlog31x0xx} show this to be the case.

A comparison of the rescaled data between the unknot and the
trefoil in figure \ref{figureFlog31-01} shows minor differences
in the rescaled pressure between these knot types.  It is not
clear that this is due to small corrections to scaling, but the
result is consistent with trefoil knots having small relative
excess rescaled pressures at short distances from the origin, 
and small deficit rescaled pressures
at longer at the surface layer, compared to the unknot. 
This pattern is also seen in figures \ref{figureFlog31-41}
and \ref{figureFlog01--52}, where more complicated knot types
are examined.  Those results suggest that more complicated 
knots have increased pressure close to the origin, and 
a reduced or a deficit of pressure near and in the surface layer. 

The results in this study can be extended to models of linear
polymers using the same approaches.  A more complex situation
may be encountered if the polymer is grafted to a hard wall.  These
are ongoing projects and the results will be forth-coming.

\vspace{1cm}
\noindent{\bf Acknowledgements:} EJJvR acknowledges financial support 
from NSERC (Canada) in the form of a Discovery Grant.

\bibliographystyle{plain}
\bibliography{References}

\end{document}